\documentclass{article}
\usepackage{amssymb}

\newcommand{\Z}{{Z \!\!\! Z}}
\newcommand{\beqn}{\begin{eqnarray}}
\newcommand{\eeqn}{\end{eqnarray}}
\newcommand{\eq}[1]{(\ref{#1})}

\newcommand{\dD}{{\mathrm D}}
\newcommand{\dual}{\mbox{}^{\ast}}

\newcommand{\cint}[1]{\!\int\nolimits^\pi_{-\pi} \!\!\!\!\dD #1}

\newcommand{\nsum}[2]{\sum_{#1 \in \Z(c_#2)}}
\newcommand{\ClosedSum}[2]{\sum_{\stackrel{#1 \in \Z(c_#2)}{\delta #1 = 0}}}

\begin{document}

\title{\bf{\sc{string breaking and monopoles}}\footnote{\uppercase{P}resented by
the second author at ``\uppercase{C}onfinement \uppercase{V}'',
\uppercase{G}argano, \uppercase{I}taly, 10-14 \uppercase{S}ep. 2002.}
}

\author{
\sc{\lowercase{M.N.~Chernodub}}\footnote{\uppercase{S}upported
by \uppercase{JSPS} \uppercase{F}ellowship
\uppercase{P}01023.}$\,\,{}^{,\lowercase{a,b}}$
\lowercase{and}
\sc{\lowercase{Tsuneo~Suzuki}}\footnote{\uppercase{S}upported
by \uppercase{JSPS}
\uppercase{G}rant-in-\uppercase{A}id for \uppercase{S}cientific
\uppercase{R}esearch on \uppercase{P}riority \uppercase{A}reas
13135210.}$\,\,{}^{,\lowercase{a}}$
}
\date{\normalsize{\it{
${}^a$ ITP, Kanazawa University, Kanazawa, 920-1192, Japan \\
${}^b$ ITEP, B. Cheremushkinskaya 25, Moscow, 117259, Russia}}}


\maketitle

\begin{abstract}
The string breaking is discussed in $U(1)^{N-1}$
Abelian effective theories of QCD. When a screening is expected,
the static potential shows a flattening in the long-range region
and a linear behavior in the intermediate region. We show why the
screening is better observed in the Polyakov-loop correlators than
in the Wilson loops. The breaking of the adjoint string is
explained without the Z(N) picture.
\end{abstract}

\vskip 3mm

The idea~\cite{'tHooft:1981ht} of monopole condensation has been
shown to be successful in $SU(N)$ QCD. Abelian and monopole
components of the string tension are dominant in the
long-range region of quenched QCD~\cite{suzuki90}. Although the
't~Hooft scenario of color confinement seems correct in many
respects, there remains an unsolved serious problem even in
quenched QCD and also in full QCD. It is the screening-confinement
problem extensively discussed in recent
publications~\cite{StringBreaking}. Following
Ref.~\cite{SuzukiChernodub} we show how the screening and
confinement problem is solved qualitatively in the framework of
$U(1)^{N-1}$  Abelian dynamics. We find the effect of dynamical
charged particles can be described in terms of integer  electric
currents.

Let us first consider the case without a dynamical char\-ged
particle. For simplicity we study the case $SU(2)$ QCD  $\to
U(1)$. The corresponding Abelian projected model is described by
the compact QED with a modified action. In the formalism of
differential forms the compact QED reads as:
\beqn
Z_1 =
\cint{\theta} \!\!\!\!\!\nsum{n}{2} \!\!\!\!\!
e^{ - \frac{1}{4\pi^2}({\mathrm d} \theta + 2 \pi n, \Delta D
({\mathrm d} \theta + 2 \pi n)) + i (Q J,\theta)} =
\!\!\!\!\! \sum_{\stackrel{\sigma \in \Z(c_2)}{\delta\sigma=QJ}}
\!\!\!\!\! e^{-\pi^2(\sigma, (\Delta D)^{-1}\sigma)},
\label{Z1}
\eeqn
where the operator $D$ is a general differential operator, $Q$ is
a charge of an external source and the current $J$ takes $\pm 1$
along the Wilson loop. $\Delta$  is the  Laplacian on the lattice.
In the second part of this equation -- which is written in the string
representation -- the summation goes over the surfaces spanned
($|Q|$ times) on the external current $J$.

In the infrared limit the operator $D$ can be approximated
with a good accuracy by
Coulomb+self+nearest neighbor terms~\cite{nakamura}:
$D=\beta\Delta^{-1}+\alpha+\gamma\Delta$, where $\beta$, $\alpha$
and $\gamma$ are renormalized coupling constants of the monopole
action, $\beta\gg\alpha, \gamma$. Then
the Wilson loop can be estimated from Eq.\eq{Z1}:
\beqn
{\langle W_Q(R,T) \rangle}= {\mathrm{const.}}\,
e^{ - \kappa\, |Q| \, RT + \cdots}\,,
\label{string:mon}
\eeqn
where $RT$ is the area of the minimal surface spanned on the
contour $J$. There are $|Q|$ such surfaces which must be parallel
to each other~\cite{Faber} to maximize the contribution for $|Q|
\geqslant 2$. This explains linearity in $|Q|$ in
Eq.~\eq{string:mon}.
Since $\beta\gg\alpha, \gamma$,  the string
tension is $\kappa \approx \pi^2 \slash \beta$.

Now let us introduce a dynamical charged particle. Consider a
charged scalar Higgs field in the London limit as a simplest
example. The radial part $\rho_x$ of the scalar Higgs field
$\Phi_x = \rho_x \, e^{i \vartheta_x}$ is frozen and the dynamical
variable is the compact Higgs phase $\vartheta \in [-\pi,\pi)$
which carries the electric charge $q$. The Higgs field action in
the Villain representation~is:
\beqn
Z_2[\theta] = \! \cint{\vartheta} \!\!\!\!
\nsum{l}{1} \!\!\!\!\!\!
e^{ -({\mathrm d} \vartheta + q \theta + 2 \pi l, G
({\mathrm d} \vartheta + q \theta + 2 \pi l))}
= \!\!\!\! \ClosedSum{j}{1} \!\!\!\!
e^{-(j, (4G)^{-1} j)+i(qj,\theta)}\,,
\label{Z4}
\eeqn
where $G$ is a local operator. The integration of the phase of
the Higgs field $\vartheta$ is represented as a weighted sum
of the Wilson loop over the closed charged
currents~\cite{EihornSavit} in the second part of Eq.~\eq{Z4}.
Hence the Villain type compact QED with the charged scalar field is
\beqn
Z_3 \!\! = \!\!\ClosedSum{j}{1}
\!\!\! e^{-(j, (4G)^{-1} j)}\cint{\theta}
\!\!\!\!\nsum{n}{2} \!\!\!\!
e^{ - \frac{1}{4\pi^2}({\mathrm d} \theta + 2 \pi n, \Delta D
({\mathrm d} \theta + 2 \pi n)) +i(QJ+qj,\theta)}.
\label{Z5}
\eeqn
This expression can be reduced further to
the string-electric current model:
\beqn
Z_4 = \ClosedSum{j}{1}
\sum_{\stackrel{\sigma \in \Z(c_2)}{\delta\sigma=QJ+qj}}
\hspace{-.3cm}e^{-(j,(4G)^{-1}j)
-\pi^2(\sigma, (\Delta D)^{-1}(\sigma))}\,,
\label{Z6}
\eeqn
where $\dual n\equiv\Delta^{-1}\delta\dual s$ and
$s$ is a surface spanned $q$ times on the dynamical current $j$,
{\it i.e.}, $\delta s\equiv qj$.

{}From Eq.(\ref{Z6}), we can see how the
screening-con\-fi\-ne\-ment problem is solved. If $Q/q \notin \Z$,
 the static potential can not be screened completely. For
example, in the case of $Q=1$ and $q=2$, the leading string
tension is equivalent to the string tension without presence of
the dynamical charges as in (\ref{string:mon}). In this case,
$j=0$ in the sum over $j$ gives the leading term while other terms
coming from non-zero $j$ show stronger damping.

On the other hand, if $Q/q \equiv N \in \Z$, then the expectation
value of the $R\times T$ Wilson loop is expanded as a perimeter
term ($\propto (R+T)$) given by $QJ+qj=0$ contribution and the
area-law term:
\beqn
{\langle W_Q(R, T) \rangle} = c_0 \, e^{-m N^2 (R+T)}
+ c_1 \, e^{- \kappa q R T - m (N-1)^2 (R+T)}  + \cdots\,,
\label{string:all}
\eeqn
corresponding to the screening currents $j = - N J$ and $j = (- N
+ 1) J$, respectively. The string breaking is seen at large
distances while the area-law behavior is observed in the
intermediate region.

Next, consider two Polyakov loops
separated by a distance $R$ corresponding to a pair of static
quark and anti-quark. The two leading terms in the average are:
$
{\langle P_Q(0)P_{\bar Q}(R) \rangle} = d_0 \, e^{-m N^2 T}
+ d_1 \, e^{- \kappa q R T - m (N-1)^2 T}+ \cdots\,.
$
A comparison of this expression with Eq.\eq{string:all} gives that
the area-law terms are the same for both averages. However, the
perimeter terms are different: for the Wilson loop the perimeter
term contains additional suppression factor $\sim e^{- m (2N - 1)
R}$ with respect to the area term. Consequently, at sufficiently
large separations between the sources the perimeter term in the
Wilson loop average may not be found numerically. Thus, in
agreement with numerical results~\cite{StringBreaking},
the string breaking may not be observed in the Wilson loop even if
the breaking is seen in the Polyakov line~correlator.

Similar ideas can be applied to the adjoint Wilson loop
screening in the pure $SU(2)$ gluodynamics as well as to the
screening of fundamental charges in the full QCD. The
relevant details can be found in Ref.~\cite{SuzukiChernodub}.

\end{document}